# Demystifying speckle field quantitative phase microscopy

Azeem Ahmad[1,†,*], Nikhil Jayakumar[1,†], and Balpreet Singh Ahluwalia[1,2,*]

[1]Department of Physics and Technology, UiT The Arctic University of Norway, Tromsø 9037, Norway
[2]Department of Clinical Science, Intervention and Technology Karolinska Institute, and Center for Fetal Medicine, Karolinska University Hospital, Stockholm 17177, Sweden
*Corresponding Email: ahmadazeem870@gmail.com and balpreet.singh.ahluwalia@uit.no

[†]These authors contributed equally to this work.

**Abstract:** Quantitative phase microscopy (QPM) has found significant applications in the field of biomedical imaging which works on the principle of interferometry. The theory behind achieving interference in QPM with conventional light sources such as white light and lasers is very well developed. Recently, the use of dynamic speckle illumination (DSI) in QPM has attracted attention due to its advantages over conventional light sources such as high spatial phase sensitivity, single shot, scalable field of view (FOV) and resolution. However, the understanding behind obtaining interference fringes in QPM with DSI has not been convincingly covered previously. This imposes a constraint on obtaining interference fringes in QPM using DSI and limits its widespread penetration in the field of biomedical imaging. The present article provides the basic understanding of DSI through both simulation and experiments that is essential to build interference optical microscopy systems such as QPM, digital holographic microscopy and optical coherence tomography. Using the developed theory of DSI we demonstrate its capabilities of using non-identical objective lenses in both arms of the interference microscopy without degrading the interference fringe contrast and providing the flexibility to use user-defined microscope objective lens. It is also demonstrated that the interference fringes are not washed out over a large range of optical path difference (OPD) between the object and the reference arm providing competitive edge over low temporal coherence light sources. The theory and explanation developed here would enable wider penetration of DSI based QPM for applications in biology and material sciences.

1. Introduction

During the last few decades, speckle interferometry has been widely used in the fields of electronic speckle pattern interferometry (ESPI) and speckle shearing interferometry (SSI) to measure the in-plane and out of plane displacement and vibration of the objects with optically rough surfaces [1, 2]. Speckle field illumination has also been implemented in various other fields of optical imaging such as structured illumination microscopy [3], laser speckle contrast imaging (LSCI) [4], near-field Fourier ptychography [5], rotating coherent scattering (ROCS) microscopy [6] etc. Recently, temporally varying or dynamic speckle illumination (DSI), which effectively reduces/removes the speckle noise from the images, has been employed in optical profilometry and quantitative phase microscopy (QPM) of biological specimens [7-14]. This type of illumination has low spatial and high temporal coherence properties unlike conventional light sources such as halogen lamp, LEDs and lasers. It is recently demonstrated that this light source has several advantages such as high spatial phase sensitivity, single shot, scalable field of view (FOV) and resolution and high space bandwidth product as illustrated in Fig.1 [15].

In the early days of interferometry, thermal light sources/white light (WL) spatially filtered with the help of pinhole are being employed due to absence of pure monochromatic light sources like lasers. The pinhole improves the spatial coherence of the light source at the expense of huge intensity loss. Thus, restricts the practical applications of optical interferometry techniques such as QPM and DHM. In addition, WL source has large spectral bandwidth, i.e., composition of large number of monochromatic spectral components, which confines the interference fringe in a limited interference field of view (iFOV) of the camera Fig. 1(a). Moreover, it also restricts the implementation of QPM only for a fixed objective lens as identical objective lenses are required in both object and reference arm to match the optical path length within the temporal coherence (TC) length (1 – 2 µm) of the light source. The iFOV can be increased by employing either a narrow bandpass light emitting diode or inserting a spectral filter in the WL beam path, also called filtered white light (FWL), at the cost of further intensity loss (see Fig. 1(b)). On the contrary, narrowband lasers overcome the restriction of the limited iFOV as in case of WL/FWL (Fig. 1(c)). However, it degrades the image quality due to the presence of coherent noise and parasitic fringes which are generated due to the large TC length of the lasers. As a consequence, it reduces the spatial phase sensitivity and height measurement accuracy of the system. Contrary to the conventional light sources, DSI has high TC length almost equal to the TC length of the parent laser light source and low spatial coherence (SC) length depending on the source size. High TC and low SC length of DSI helps to achieve coherent noise free interference pattern over the whole camera FOV unlike conventional light sources (Fig. 1(d)).

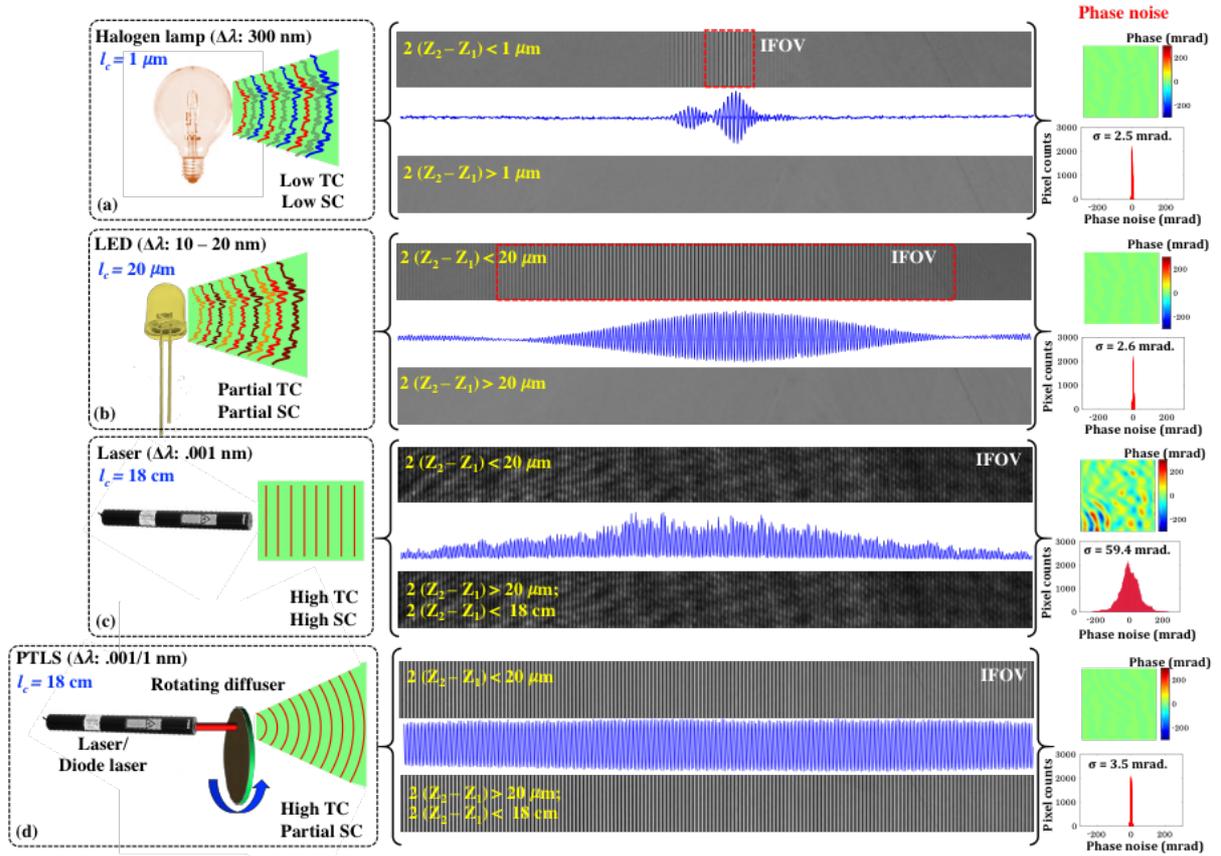

**Fig. 1. Comparison of conventional light sources.** a) white light using halogen lamp, b) light emitting diode or filtered white light, c) laser and d) DSI. The comparison in terms of the extent of iFOV, quality of the interference fringes and spatial phase sensitivity. $Z_1$ and $Z_2$ demonstrate the optical path lengths of the object and the reference arms of the interferometer. $2(Z_2 - Z_1)$ is the optical path difference between the object and the reference arm, $\Delta\lambda$ is the spectral bandwidth and $l_c$ is the temporal coherence length of the light source. The line profiles along the full FOV are depicted in blue color.

Despite its edge over the conventional light sources, dynamic speckle illumination quantitative phase microscopy (DSI-QPM) is limited, possibly due to lack of in-depth understanding. In DSI-QPM, generation of correlated speckle field is mandatory to form interference pattern when the reference and the object arm speckle fields are superimposed to each other [12]. However, there exists some common misconceptions about the experimental conditions under which the interference fringes in DSI-QPM system are obtained [11-13], such as:

a) Use of identical objective lenses in both the object and the reference arms of the interference microscopy system to generate correlated speckle fields,
b) Field of view (FOV) fixed and dependent on pre-chosen imaging objective lens,
c) Short optical path difference (OPD) adjustment range between the object and the reference arm of the QPM system to observe interference pattern at the detector,
d) Dependency on diffraction grating in the reference arm to obtain high density fringes over a large field of view (FOV) of the camera.

These doubts also bring the following questions into the discussion: a) What are the underlying experimental conditions behind the formation of interference fringes in DSI-QPM system? b) Does DSI-QPM follows the laws of conventional interferometry in terms of interference fringe's shape of the resultant pattern?

The lack of in-depth theoretical understanding behind the formation of interference fringes in DSI-QPM system restricts the widespread penetration of DSI in the field of phase microscopy. In addition, the underlying experimental conditions to obtain interference fringes in DSI-QPM have not been convincingly covered previously. The present article provides the theoretical framework and systematic simulation and experimental studies behind the formation of interference pattern in DSI-QPM resolving previous ambiguities such as use of identical objective lens for the formation of interference fringes and creating of high-density fringes via grating. Here, both simulation and experimental studies are performed to understand the conditions required to achieve stable interference patterns in DSI-QPM. This is done by providing a theoretical framework of the superposition of correlated and uncorrelated speckle fields and its effect on the resultant. It is observed that the speckle fields being overlapped must be correlated and unshifted with respect to each other to obtain high contrast interference fringes in DSI-QPM. In the reflection geometry, this enables the use of non-identical objective lenses in the object and the reference arm of the system provided the above conditions are satisfied. Thus, it provides flexibility to the user to achieve scalable FOV and resolution in the system. Moreover, user defined fringe density (low or high) can easily be obtained in DSI-QPM system without using grating in the beam path and provide an ease to implement either single-shot or multi-shot QPM.

## 2. Theory

### 2.1. Correlated and uncorrelated speckle pattern

From the statistical theory of optical fields, the speckle fields must be statistically independent or decorrelated to reduce the speckle contrast, i.e., noise, from the images of the test specimens [16]. On the contrary, speckle interferometry requires correlated speckle fields (statistically dependent) which are being superimposed to each other to form stable speckle interferograms or specklograms [11]. If the speckle fields being superimposed with each other are not correlated (statistically independent), then the resultant intensity pattern represents unstable specklogram. The correlation between the speckle fields controls the shape of the resultant intensity patterns.

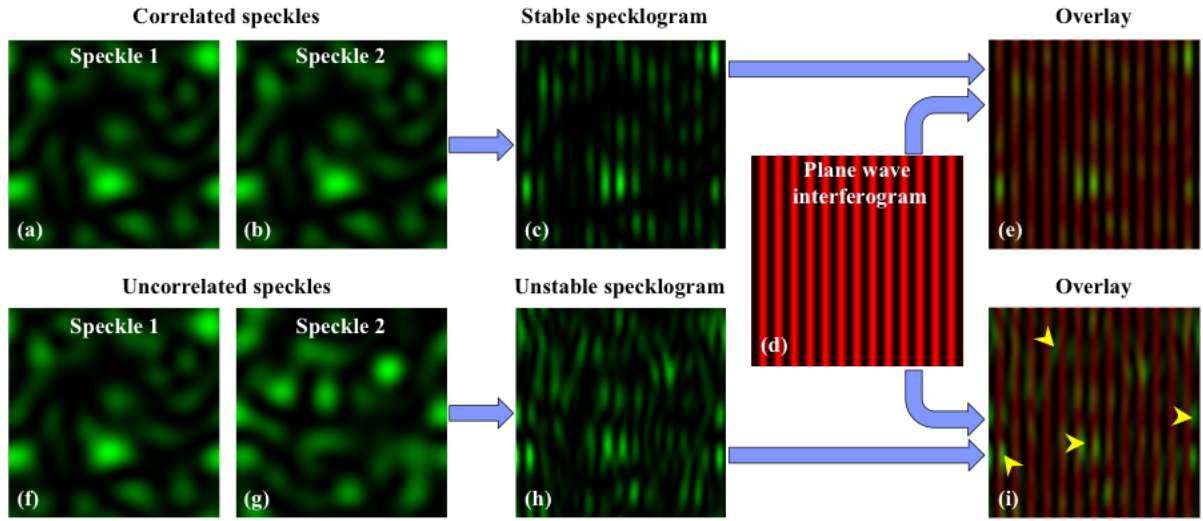

**Fig. 2. Representation of the superposition of correlated and uncorrelated speckle fields at the detector.** The green regions exhibit the speckles in the images. For the correlated case, the interference fringes formed within the speckles follow the orientation of reference fringes and also do not show misalignment in the fringe positions from speckle to speckle. This is represented as a stable specklogram. Whereas, in the uncorrelated case, the fringes present in one speckle are not aligned to the fringes present in other speckles. This leads to the formation of randomly oriented interference fringes in the specklogram. This is termed as an unstable specklogram.

Figure 2 provides a simulation study to demonstrate the interference fringes formed by correlated and uncorrelated speckle patterns. Figures 2(a) and 2(b) exhibit two correlated speckle patterns. The superposition of these speckle patterns does not deform the shape of the resultant intensity pattern and forms nicely oriented specklogram as illustrated in Fig. 2(c). It can be seen that the interference fringes do not locally follow the shape and positions of the speckles present in the field of view. On the contrary, the superposition of two decorrelated speckle fields shown in Figs. 2(f) and 2(g) does not generate nicely oriented interference fringes. It is exhibited in Fig. 2(h) that the resultant interference fringes have random orientation locally and follow the shape and position of speckles. The deformation in the shape of the resultant interference fringe patterns is exhibited more clearly by overlaying them with the reference interferogram (Fig. 2(d)) generated due to the superposition of two plane waves. The angle between the plane waves is kept equal to the angle considered during speckle fields superposition. Figures 2(e) and 2(i) illustrate the overlayed images of the stable and unstable specklograms with the reference interferogram. The fringes in the stable specklogram exactly overlap with the reference fringes. Whereas, for unstable specklogram, the fringes do not overlap with the reference fringes. Some of such locations are exhibited by yellow arrows in Fig. 2(i).

### 2.2. Speckle theory

The superposition of multiple randomly phased complex components results into an irregular pattern in 3D space called speckle pattern, which is a granular like structure [16]. Mathematically, the resultant phasor $A(x, y, z)$ at a single point in space–time can be represented as follows [16]:

$$A(x, y, z) = \frac{1}{\sqrt{N}} \sum_{k=1}^{N} a_k e^{i\phi_k} = A e^{j\theta} \quad (1)$$

where, N is the total number of randomly phased complex components of amplitudes $a_k/\sqrt{N}$ and the phases $\phi_k$. A and θ are the amplitude and phase of the resultant phasor, respectively.

In general, if we have two or more speckle patterns then their sum can be done either on the basis of amplitude or intensity. The speckle pattern summed on the amplitude basis does not change the statistical distribution of the amplitude and intensity both [16]. Thus, the speckle contrast of the final speckle pattern does not reduce. On the contrary, addition on the intensity basis does reduce the fluctuation or contrast of the speckle provided some degree of decorrelation exist between the speckle patterns being added [16]. If the detector integrates M1 uncorrelated or statistically independent speckles, then the response of the detector is the summation of their intensity patterns. Thus, the total intensity is given by the sum

$$I_T = \sum_{m=1}^{M} I_m = \sum_{m=1}^{M} |A_m|^2 \qquad (2)$$

The correlation between the nth and mth intensity patterns is represented by the following expression [16]

$$C_{nm} = \frac{\langle I_n I_m \rangle - \langle I_n \rangle \langle I_m \rangle}{[\langle (I_n - \langle I_n \rangle)^2 \rangle \langle (I_m - \langle I_m \rangle)^2 \rangle]^{1/2}} \qquad (3)$$

The value of correlation lies between 0 and 1. The correlation equal to 1 and 0 corresponds to the statistically dependent (unreduced speckle contrast) and independent (reduced speckle contrast) speckle pattern intensities, respectively. The contrast of the averaged image depends on the number of speckle patterns ($M_1$) being added and defined by $k = 1/\sqrt{M_1}$. Thus, sufficiently large number of uncorrelated speckle patterns must be averaged to observe the speckle free imaging. The correlation between the speckle pattern depends on several factors like amount of translation/rotation of rough surface, geometry of the rough surface and size of the speckle [16].

The speckle patterns can be classified into two categories called objective and subjective speckle. The objective speckles are obtained on the screen if a lens is not involved in the optical configuration. Whereas, the speckle patterns formed at the image plane of the optical system that incorporates a lens are subjective speckle.

The average size of the objective speckle on the observation plane is given by [16, 17]

$$d_o = \frac{\lambda z}{D} \qquad (4)$$

where, λ is the illumination wavelength, z is the distance between the rough surface and observation plane and D is the beam diameter at the rough surface. For subjective speckles, the average speckle size at the image plane is given by the following expression [17]

$$d_s = 1.22(1 + M)\lambda F \qquad (5)$$

where, M is the magnification of the optical system and F is the f-number (focal length/effective aperture) of a lens.

## 2.3. Dynamic speckle interferometry

In the paraxial wave approximation, the complex amplitude of the speckle field at a point $(x, y)$ in the observation plane is related to the complex amplitude $A(\xi, \eta)$ of the scattered wave field at a location $(\xi, \eta)$ right after the scattering surface plane by the following Fresnel diffraction integral [16]:

$$A(x,y) = \frac{e^{jkz}}{j\lambda z} e^{j\frac{k}{2z}(x^2+y^2)} \iint_{-\infty}^{\infty} A(\xi,\eta) e^{j\frac{k}{2z}(\xi^2+\eta^2)} e^{-j\frac{2\pi}{2z}(x\xi+y\eta)} d\xi d\eta \qquad (6)$$

The above Eq. (6) is simply the Fourier transform of the product of the complex amplitude right after the rough surface and a quadratic phase factor.

The complex amplitude $A(\xi,\eta)$ of the scattered wave field at the scattering surface

$$A(\xi,\eta) = A e^{j\phi(\xi,\eta)} \qquad (7)$$

where, $\phi(\xi,\eta)$ is the random phase introduced in the incident wave due to the scattering surface.

The superposition of two speckle fields, on-axis and tilted, at a particular instant of time, say $A_1(x,y)$ and $A_2(x,y)$, can be represented by the following relation:

$$H(x,y) = \left| A_1(x,y) + A_2(x,y) e^{j2\pi(f_x x + f_y y)} \right|^2 \qquad (8)$$

where, $H(x,y)$ is the intensity pattern generated due to the superposition of two speckle fields at the detector also called specklogram. $f_x$ and $f_y$ are the global spatial frequencies of the speckle field $A_2(x,y)$ along x and y axis and can be given as follows:

$$f_x = \frac{\cos(\theta_x)}{\lambda}; \quad f_y = \frac{\cos(\theta_y)}{\lambda} \qquad (9)$$

where, $\theta_x$ and $\theta_y$ are the angles of the propagation direction of the speckle field from x and y axis. The intensity patterns of the speckle fields $A_1(x,y)$ and $A_2(x,y)$ can be calculated by the following relations

$$I_1(x,y) = |A_1(x,y)|^2; \quad I_2(x,y) = |A_2(x,y)|^2 \qquad (10)$$

The specklogram is completely filled of speckle noise which can be reduced by either rotating or translating the rough surface or diffuser. The rotating diffuser generates temporally varying speckle patterns. If large numbers of temporally varying speckle fields are averaged within the integration time 'T' of the detector, then the integrated intensity can be represented as follows:

$$I_T = \frac{1}{T} \int_0^T H(t) dt \qquad (11)$$

The average (Eq. (11)) of large number of unstable and stable specklograms corresponding to different speckle fields, form a constant background and nicely oriented interferogram, respectively. Mathematically, it can be written as follows:

$$I_T = \begin{cases} \text{const or mean of } I_T & C_{12} = 0 \\ \text{Interferogram} & C_{12} = 1 \end{cases} \qquad (12)$$

where, $C_{12}$ is the correlation between the speckle fields $A_1(x,y)$ and $A_2(x,y)$ being superimposed to generate specklograms. If the speckle fields are identical, i.e., exact replica, then they exhibit the highest correlation and considered correlated to each other. On the other hand, uncorrelated speckles correspond to the non-identical speckle fields. In other words, Eq. (12) implies that if the speckle fields are uncorrelated ($C_{12} = 0$) then interference pattern will not be observed. To generate high contrast interferograms, the speckle fields must be correlated ($C_{12} = 1$) to each other. The correlation values in between 0 and 1 corresponds to the reduced fringe visibility of the resultant interferogram.

Furthermore, two identical fields can also exhibit decorrelation when one of the speckle field is shifted or translated (say by '$\Delta x; \Delta y$') with respect to the other one. The extent of the shift within which the fields are still correlated is called correlation length '$l_c$' and is decided by the speckle size (Eqs. 4 and 5). Thus, the two identical speckle fields do not form the interference pattern if the translation is greater than the correlation length/speckle size in one of the speckle fields. For identical speckle fields, Eq. (12) in terms of shift or transition can be modified as follows:

$$I_T = \begin{cases} \text{const or mean of } I_T & C_{12} = 0 & \text{or } \Delta x; \Delta y \geq l_c \\ \text{Interferogram} & C_{12} = 1 & \text{or } \Delta x; \Delta y = 0 \\ & 0 < C_{12} < 1 & \text{or } \Delta x; \Delta y < l_c \end{cases} \quad (13)$$

## 3. Simulation studies

### 3.1. The superposition of correlated and uncorrelated speckle patterns

A systematic simulation study is done to understand the interference of the correlated and uncorrelated speckle patterns and its influence on the resultant intensity patterns. We first discuss the reduction of the speckle noise from the images and then the interference of the speckle patterns both correlated and uncorrelated. In optical interferometry, the speckle patterns are called correlated or statistically dependent if they match elementwise (or pixel-wise) with each other. The uncorrelated speckle patterns do not match element-wise with each other and also called statistically independent. The correlation between the speckle patterns is calculated by employing Eq. (3).

In order to get rid of the speckle noise, a large number of uncorrelated speckle patterns must be added. There are various ways to generate statistically independent speckle patterns which are based on temporal, polarization and wavelength diversity [16]. The temporal diversity which is used throughout the present work is obtained by rotating the rough surface or diffuser. The rotating diffuser is the simplest and effective method for speckle noise reduction from the images [16].

Generation of uncorrelated speckle patterns is needed to reduce the speckle noise from the images. On the contrary, in the field of interferometry the speckle patterns must be correlated or statistically dependent to generate interference pattern due to their superposition at the detector. In the simulation, a rough object having height variations h(x,y) is created as shown in supplementary Fig. S1. The height variation is chosen in a way such that it corresponds to the phase variation uniformly distributed between – π to + π. The phase variation of the rough surface is related to its height variation by the following relation:

$$\phi(x,y) = \frac{2\pi}{\lambda} h(x,y) \quad (14)$$

The transmittance of the diffuser is given by the following expression:

$$T(x,y) = T_0(x,y) \exp\left(i \frac{2\pi}{\lambda} h(x,y)\right) \quad (15)$$

where, $T_0(x,y)$ is the transmission coefficient of the diffuser, $\lambda$ is the wavelength of the incident light.

Let us consider a circular light beam of unity amplitude and diameter D hits the diffused surface. Numerically, it can be done by multiplying the transmittance given in Eq. (15) with a circular mask (diameter: D) having 1s inside the circle and 0s outside the circle as illustrated in supplementary Fig. S2. The speckle pattern is then generated by performing its Fourier transform (Eq. 6) and multiplying element wise by the complex conjugate [18].

First, two uncorrelated speckle fields having angle 'θ' are generated by simulating two different diffusers and further superimposed to study their effect on the resultant intensity pattern. This situation is analogous to the optical interference of two random speckle fields. The ray diagram of the configuration is depicted in Fig. 3(a). It can be seen from Fig. 3(a) and 3(b) that the global phase fronts (black dotted lines) of the two speckle fields (say R and O) have spherical shape before the tube lens (TL). The TL collimates both the speckle fields and overlaps them at plane P. In Fig. 3(a), both the beams have same curvature (say $C_1$) and different speckle fields ($S_1$ and $S_2$) before TL. It is observed that the resultant intensity pattern does not have any fringe like pattern as illustrated in Fig. 3(a). Thus, statistically independent fields do not form nicely oriented interference pattern. On the contrary, the superposition of two correlated speckle fields form a nicely modulated intensity pattern at plane P as depicted in Fig. 3(b). Note that the speckle noise does not affect the global shape of the modulated intensity pattern.

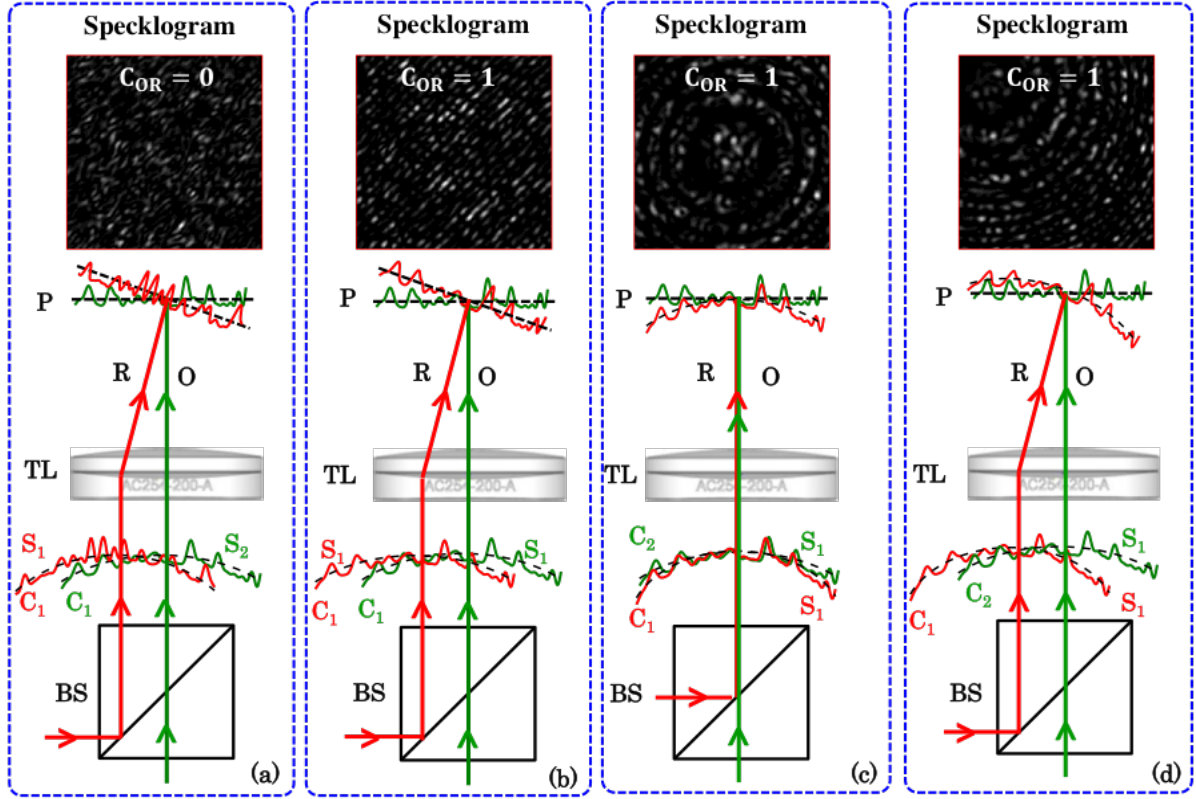

**Fig. 3. Optical configurations to superimpose the correlated and uncorrelated speckle fields.** (a) and (b) represent the superposition of two uncorrelated and correlated speckle fields in off-axis configuration, respectively. The global phase fronts (black dotted lines) are assumed to be plane wavefronts; on-axis and tilted as shown in Fig. 3(a) and 3(b). (c) and (d) represent the superposition of two correlated speckle fields in the on-axis and off-axis configuration, respectively. The global phase fronts of the speckle fields are spherical and plane. The shape of the global phase fronts (black dotted lines) does not the affect the quality/contrast of the specklograms. It only changes the shape of the fringes in the resultant pattern.

Next, we investigated the case when two correlated speckle fields (S1) with different global spherical phase fronts ($C_1$ and $C_2$ before TL) are superimposed at the focal plane of TL as shown in Fig. 3(c). This is an on-axis superposition of two speckle fields. TL collimates one of the speckle fields having curvature $C_2$, whereas second speckle field has slight curvature after TL. This generates specklogram with spherical fringes centred at the origin as shown in Fig 3(c). Figure 3(d) illustrates the off-axis superposition of two correlated speckle fields with global phase fronts $C_1$ and $C_2$. The beam R meets the beam O at an angle as depicted in Fig. 3(d). The superposition of these two speckle fields generates off-centred circular fringes in the specklogram as shown in Fig. 3(d). This concludes that the shape of the global phase fronts does not reduce the quality of the specklogram as long as the speckle patterns are correlated. Thus, speckle interferometry follows the laws of conventional interferometry in terms of fringe shapes provided the fields are correlated.

It is therefore evident that the correlated speckle fields can generate fringe like pattern in the resultant intensity at the cost of high speckle noise as shown in Fig 3. The speckle noise can be reduced by averaging large number of specklograms generated corresponding to statistically independent speckle fields ($S_1$, $S_2$, …… $S_n$). Statistically independent speckle fields are generated by simulating a rotating diffuser. First, a rough surface is generated to mimic the static diffuser in the simulation (Fig. 4(a)). The rough surface is multiplied by a circular binary mask. The position of the opening in the mask (partially filled green circle) is kept at a distance R from the centre of the diffuser as shown in Fig. 4(a). The opening in the mask is then rotated along the circular path depicted in red color. The angular position of the mask's opening is varied from 0° – 360° in a step of 1°. This generates statistically independent speckle fields corresponding to different rotation angles of the opening. Figure 4(b) illustrates the statistically independent speckle patterns (contrast = 0.98) corresponding to the rotation angles '$\theta_{RD}$' from 0° to 360° in a step of 1°. The average of these statistically independent speckle patterns is shown in Fig. 4(b) which is having very low contrast equal to 0.13.

Since it is well known from Fourier transform's shifting property that the non-centric position of the opening affects the resultant speckle field by introducing an extra phase factor in the calculation [18]. Therefore, here before taking the Fourier transform of the simulated rough surface to generate speckle pattern, the region of the diffuser falling in the opening region of the mask is shifted to the center of the diffuser. This is done to mimic the experimental condition in which the light beam is aligned to the optical components of the phase microscope and diffuser is rotating.

Figures 4(c) and 4(e) illustrate the specklograms corresponding to the rotation angles '$\theta_{RD}$' from 0° to 360° in a step of 1° for the correlated and uncorrelated speckle fields, respectively. The global phase fronts are considered plane for both the speckle fields (R and O) being superimposed to generate specklogram. The average images of 360 specklograms corresponding to correlated and uncorrelated speckle fields are also shown in Figs. 4(c) and 4(e), respectively. The average of the specklograms formed due to correlated speckle fields, generates a super clean interferogram. For the uncorrelated case, fringe pattern is not formed as depicted in Fig. 4(c). Next, two speckle fields having global plane and spherical wavefronts are superimposed to generate corresponding specklogram. In the case of correlated speckle fields, the average of 360 specklogram generates clean and nicely oriented off-centred circular fringes as shown in Fig 4(e). The average image with approximately uniform intensity is generated for uncorrelated speckle fields as illustrated in Fig. 4(f). Thus, to generate high contrast and clean interference fringes in speckle interferometry setups, the fields must be correlated to each other. In addition, the formation of interference pattern does not depend on the shape of the global phase fronts.

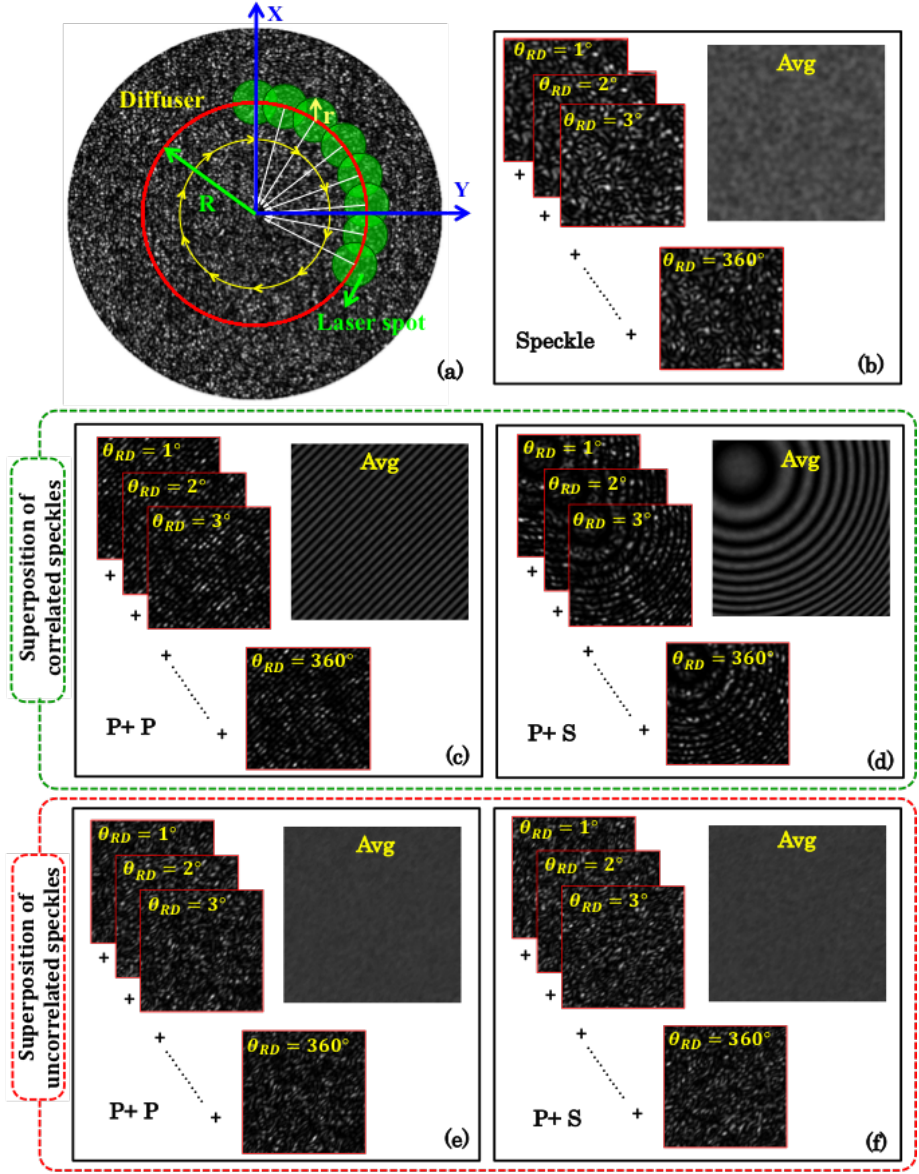

**Fig. 4. Correlation study between correlated and uncorrelated speckle fields.** (a) Mimicking of the rotating diffuser in the simulation. Green color filled circles represent the laser spot as a function of different rotation angles θ$_{RD}$. Red color solid circle is the circular path (radius R) of the laser spot (radius r) at the diffuser. (b) Represents the series of statistically independent speckle patterns corresponding to different rotation angles $\theta_{RD}$ from 0° to 360° in a step of 1° and their average image. (c) and (d) illustrate the specklograms and averaged interferograms generated due to the superposition of correlated object and reference speckle fields for plane and spherical global phase fronts, respectively. (e) and (f) exhibit the resultant intensity patterns generated due to the superposition of uncorrelated object and reference speckle fields for plane and spherical global phase fronts, respectively. P and S are defined as; P: Plane phase fronts and S: Spherical phase front.

### 3.2. Superposition of the shifted speckle fields

In this section, the effect of the superposition of a speckle field with its shifted version on the resultant intensity is systematically studied. This situation is analogous to the experimental condition where both the speckle fields (object and reference) originated from the same source do not exactly overlap to each other due to slight misalignment in the optical setup. The details about the experimental conditions are discussed in section 4. This study is conducted with the speckle fields having two different speckle sizes of 15 pixel and 30 pixel.

First, a speckle field is superimposed with its shifted version to generate specklogram for both speckle sizes of 15 pixel and 30 pixel. The shifted version is obtained from the copy of the same field. This is done by translating the speckle field along x and y direction in a step of say $\Delta x$ and $\Delta y$. The value of $\Delta x$ and $\Delta y$ is varied from zero to the average size of the speckle. Figures 5(a) – 5(c) illustrate the case of superposition of identical unshifted speckle fields at plane $P_1$. The ray diagram of the overlap of identical speckle fields (R and O) at plane P1 is shown in Fig. 5(a). The specklograms generated due to the superposition of the identical speckle fields of speckle sizes of 15 pixel and 30 pixel are depicted in the upper row of Figs. 5(b) and 5(c), respectively. The specklograms exhibited in Figs. 5(b) and 5(c) correspond to the rotation angle '$\theta_{RD}$' equal to zero. The rotation angles '$\theta_{RD}$' is defined in the previous section and varied from 0° to 360° in a step of 1° to generate statistically independent specklogram. The average images are then generated by taking the average of 360 specklograms corresponding to both speckle sizes of 15 pixel and 30 pixel and illustrated in Figs. 5(b) and 5(c). It can be visualized from the average images that the resultant modulated intensity pattern is not affected by the size of the speckle in the case of unshifted speckle fields.

One of the speckle fields is then shifted by 15 pixel and 30 pixel corresponding to the speckle sizes of 15 pixel and 30 pixel, respectively. These shifted speckle fields are then superimposed with the original/unshifted copy of the speckle field to generate specklograms as shown in Figs. 5(e) and 5(f). It can be seen from the specklograms shown in Figs. 5(e) and 5(f) that the local fringe orientations follow the shape and position of the speckles, i.e., the fringes are not nicely oriented along a particular direction as explained in the previous section. For the fringe width greater than the size of the speckle (15 pixel), the local fringes are not visible in the resultant intensity pattern. In the case of speckle size of 30 pixel, the local fringes are observed due to the smaller fringe width than the speckle size as depicted in Fig. 5(f). The corresponding average images, obtained after averaging 360 specklograms, are illustrated in Figs. 5(d) and 5(f). Thus, in order to generate interference fringes the fields being superimposed should not be shifted greater than the average size of speckles. In the simulation, high to moderate contrast (i.e., 0.9 to 0.5) interference fringes are only observed until the shift in one of the speckle field is less than the half of the speckle size.

To understand this, one of the speckle fields is shifted sequentially from 0 to the average speckle size in a step of 1 pixel and corresponding specklograms and average images are generated. The values of the correlation between the shifted speckle patterns and the original speckle pattern are then calculated using Eq. (3) and plotted as a function of shift. The corresponding average images for different shifts are exhibited in Supplementary Figs. S3 and S4. Figures 5(g) and 5(i) illustrate the correlation maps as a function of shift corresponding to the speckle sizes of 15 pixel and 30 pixel. It can be seen from the correlation map of speckle size of 15 pixel (Fig. 5(g)), the value of the correlation drops down to approximately zero for the shift equal to the speckle size. On the contrary, the correlation never drops to zero in the case of speckle size of 30 pixel as shown in Fig. 5(i). Thus, speckle fields having larger speckle size always have some correlation under shift.

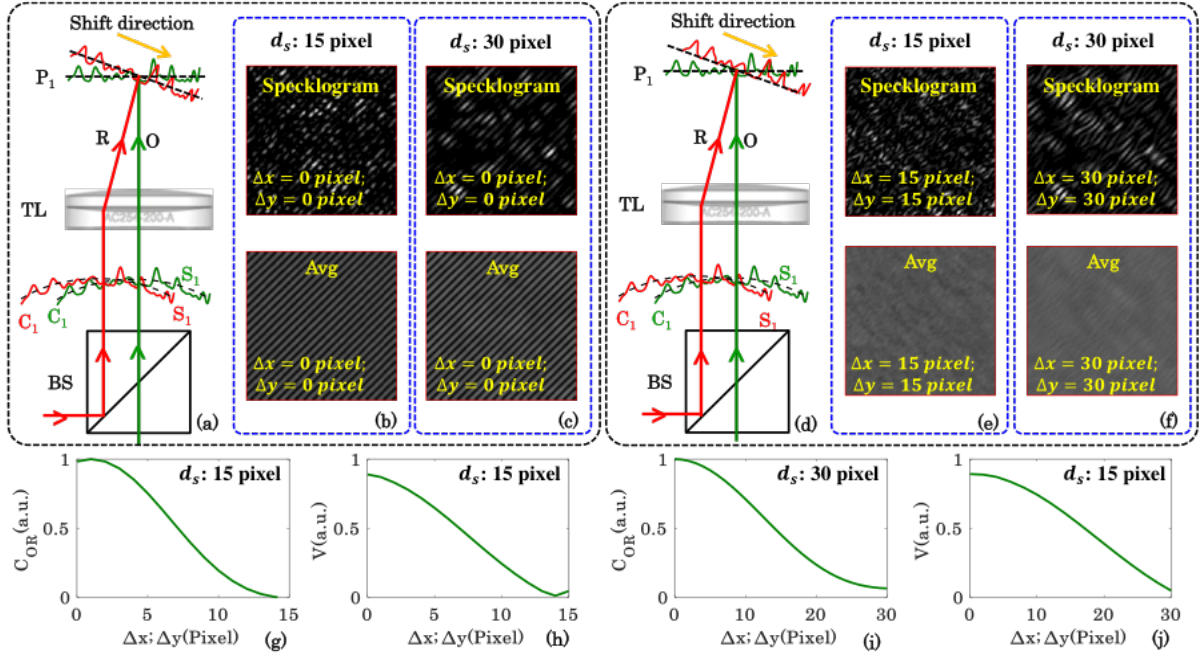

**Fig. 5. Correlation study under the shift of one speckle fields and speckle size.** The correlation between the identical speckle fields as a function of shift of one of the speckle fields being superimposed to generate resultant intensity pattern. (a) – (c) Represent the optical configuration, specklogram and averaged interferogram corresponding to the superposition of unshifted correlated speckle fields of speckle sizes 15 pixel and 30 pixel, respectively. (d) – (f) Represent the optical configuration, specklogram and averaged interferogram corresponding to the superposition of shifted correlated speckle fields of speckle sizes of 15 pixel and 30 pixel, respectively. (g) and (i) Illustrate the variation of correlation values between the speckle fields as a function of the shift in a step of 1 pixel corresponding to speckle fields of speckle sizes of 15 pixel and 30 pixel. (h) and (j) Corresponding fringe visibility plot obtained from the averaged images/interferograms.

To further confirm this, the visibilities of the average images corresponding to every shift are then calculated and plotted as depicted in Figs. 5(h) and 5(j). The details to calculate the visibility of an interferogram can be found in ref. [19]. The visibility curves also follow the similar trend as the correlation plot. However, the visibility of the interferograms does not start from 1 even for the perfectly correlated case. The slight deviation could be due to the insufficient averaging of statistically independent speckle fields being used to generate specklograms. It is also observed that average images (interferograms) have slight local variations in the fringe modulation and can be clearly seen in the supplementary Fig. S5.

### 3.3. Experimental setup

The experimental scheme of laser speckle or DSI based QPM system is illustrated in Fig. 6. The optical configuration is a Linnik interference microscopy system. The laser light beams coming from Cobolt Flamenco™ ($\lambda = 660$ nm; $\Delta\lambda = 0.001$ pm) illuminated the rotating diffuser (RD) with a beam (spot size at diffuser plane ~ 1 mm). The RD generated temporally varying statistically independent speckle fields, which added on the basis of intensity. The speed of the RD also plays an important role which controls the spatial coherence properties of the output field [16]. The scattered photons at the output of RD are directly coupled into a multi-mode fiber (Thorlabs: part # M35L01) using 20×/0.45NA objective lens as shown in Fig. 6. The core diameter of multi-mode fiber (MMF) is 1 mm. The RD followed by MMF generated uniform illumination, i.e., speckle free field, at the output port of MMF, which acts as an extended purely monochromatic light source named DSI. Before hitting the RD, the laser beam has high spatial and high temporal coherence properties. The RD and MMF significantly changed the spatial coherence properties of the light field with minimal effect on the temporal coherence properties. Thus, the output of MMF has low spatial and high temporal coherence properties.

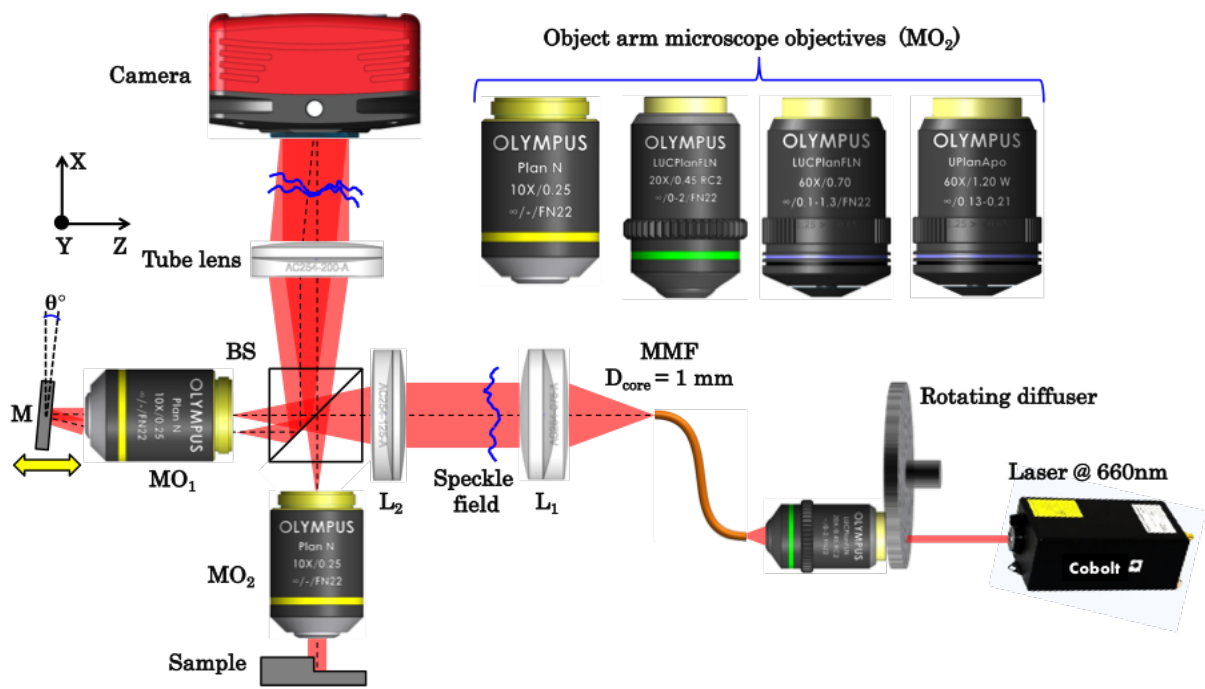

**Fig. 6. Schematic diagram of the DSI based QPM system.** $MO_{1-2}$: Microscope objectives; BS: Beam splitters; $L_{1-2}$: Lenses; MMF: Multi-mode fiber; M: Mirror and CMOS: Complementary metal oxide semiconductor camera.

The output of MMF is attached to the input port of the interference microscopy system as depicted in Fig. 6. The speckle field at the output of MMF is nearly collimated using lens $L_1$(fl: 75 mm). The second lens $L_2$ (fl: 125 mm) focused the light beam at the back focal plane of the microscope objective $MO_2$ to achieve nearly uniform illumination at the sample plane. The beam splitter BS split the input speckle field into two; one is directed towards the sample (S) and the other one towards the reference mirror. The light beams reflected from sample and mirror are recombined and overlapped at the detector using BS and a tube lens (TL).

It is highlighted in the simulation work that to obtain interference pattern the speckle fields coming from the object and reference arm must be correlated and should not be shifted more than the speckle size. If any one of the situations is not satisfied, then the interference pattern will not be observed. Thus, slight misalignment in the optical setup can completely wash out the interference pattern at the detector even though the DSI has high temporal coherence properties. The alignment of object and reference speckle fields is done with the help the reference mirror. The reference mirror is attached to the three axis precision controls stage: one translation and two angular motions. The translation stage adjusted the separation between $MO_1$ and reference mirror and tries to bring the mirror within the depth of field of $MO_1$. The kinematic mount (Thorlabs: part # KMS/M) is used to control the angle of the reference mirror along two angular directions. The translation and angular motions in the reference mirror helped to obtain correlated object and reference speckle fields at the detector. The translation of the reference mirror controlled the correlation between the speckle fields and helped to obtain high contrast fringes. The angular motion of reference mirror controlled the angle between the object and reference field or the fringe density in the camera FOV. When the angular position of the mirror is changed, the decorrelation between the speckle fields took place due to the shift in the superimposed speckle fields. As mentioned in the section 3.2, the interference fringes can be completely washout if the shift between the speckle fields is greater than or equal to the speckle size. The shift between the speckle fields is then adjusted using the translation stage to get maximum correlation or fringe contrast. Thus, three axis precision controlled the position and the orientation of the mirror and helped to obtain high contrast specklogram with variable fringe density. Due to the high TC and low SC properties of DSI, good quality interferograms are generated with variable fringe density over the entire camera FOV provided the fields are correlated at the detector.

The reference mirror and $MO_1$ is attached to a motorized stage (Thorlabs: part # MTS50-Z8) of the translation range of 50 mm. The combined motion of the mirror and $MO_1$ unit adjusted the OPD between the object arm and reference arm. This way the curvature of the interference fringes is controlled while employing different objective lenses in the object arm as illustrated in Supplementary Fig. S6.

## 4. Experimental results

### 4.1. Superposition of correlated and uncorrelated speckle fields

First, an experimental study of the simulation work presented in section 3.1 and 3.2 is done. To perform this study, the superposition of correlated and uncorrelated speckle fields at the detector is done experimentally to understand their effect on the resultant intensity pattern. In the object and reference arms 60×/0.7NA and 10×/0.25NA objective lenses are used to generate object and reference speckle fields, respectively. A piece of silicon wafer is used in both the object and the reference arms of QPM to match the reflected intensity of both the beams. The correlated object and reference speckle fields are generated when Si-wafer (in both object and reference arm) is within the depth of field of the objective lenses. In addition, precise alignment of both the object and the reference speckle fields within the correlation length is also required to generate correlated speckle fields at the detector.

Figures 7(a) and 7(b) represent the correlated object and reference speckle fields recorded by the camera by sequentially blocking the reference and the object arm of the interferometer. The diffuser is kept stationary to generate static speckle patterns. The zoomed views of the small regions of both object and reference speckle patterns marked with green, red and blue solid colour boxes are also exhibited to show the similarity between both the speckle patterns. The normalized correlation between both the patterns is calculated to be equal to 0.68 and corresponding normalized correlation map is given in the Supplementary Fig. S7(a). Next, both the object and the reference arms of the QPM are unblocked to record the specklogram as illustrated in Fig. 7(c). The diffuser is then rotated to remove the speckle noise from the specklogram as explained in the section 3.3. The rotating diffuser didn't affect the interference fringes and generated nicely oriented coherent noise free interferograms as depicted in Fig. 7(d).

Further, the reference Si-wafer is translated and tilted to generate decorrelation between the object and reference speckle fields. The decorrelated object and reference speckle fields are illustrated in Figs. 7(e) and 7(f), respectively. The normalized correlation map is depicted in Supplementary Fig. S7(b) and its value is found to be equal to 0.03. It can be seen that the correlation between the speckle fields dropped down by a significant amount compared to the correlated ones. The zoomed views of the regions marked with green, red and blue colour boxes corresponding to the object and the reference speckle fields are presented in second and third rows below Figs. 7(e) and 7(f). It can be clearly seen in the zoomed views that the object and the reference fields do not match with each other. The corresponding specklogram is illustrated in Fig. 7(g). It can be visualized that the specklogram generated due to the superposition of uncorrelated speckle fields is not contained fringe like pattern. The diffuser is then rotated to generate temporally varying speckle fields which leads to the formation of uniform intensity (i.e., without interference fringes) over the camera FOV. These experimental findings are found to be in a good agreement with the simulation results presented in sections 3.1 and 3.2.

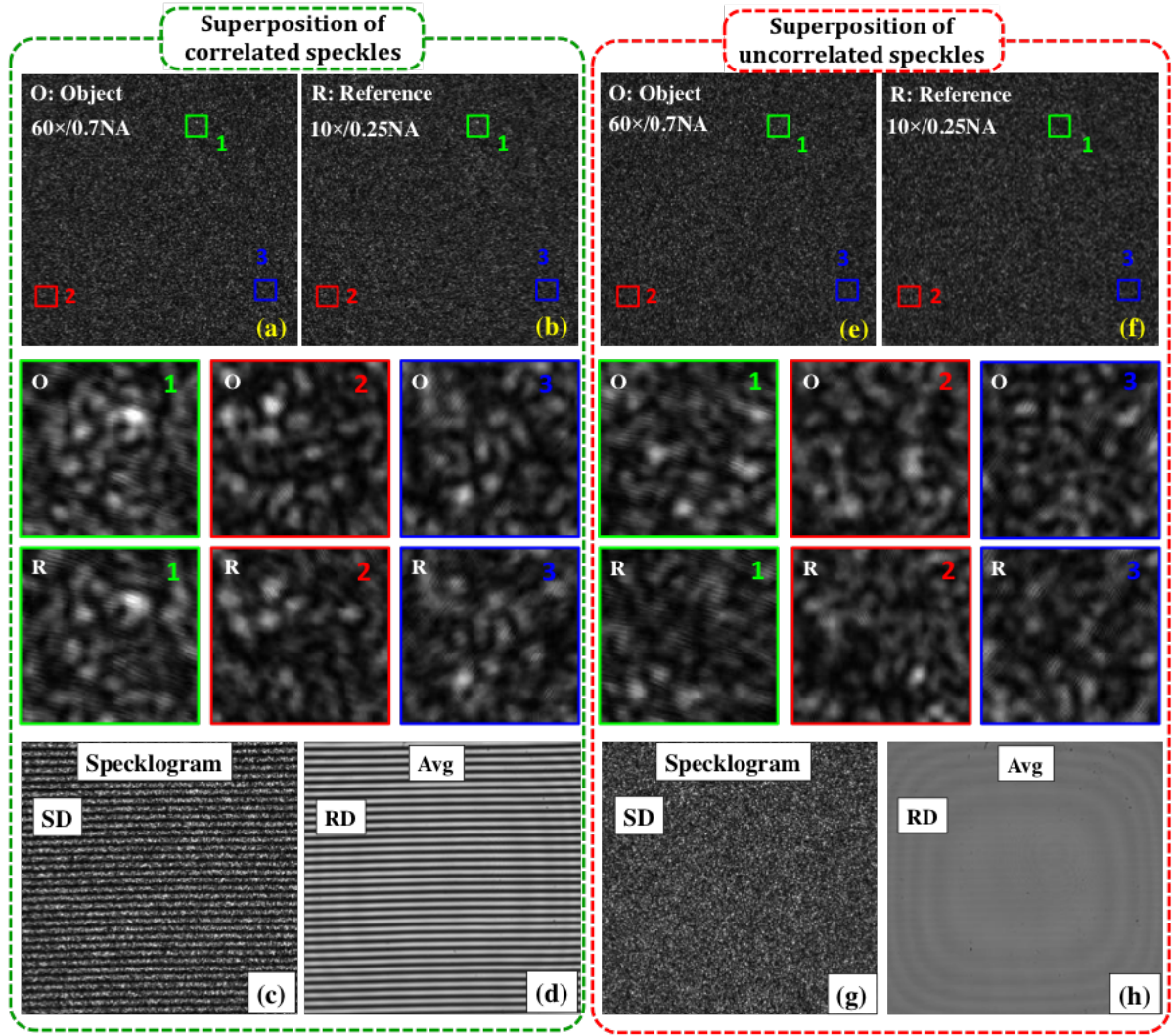

**Fig. 7. Superposition of the correlated and uncorrelated object and reference speckle fields.** The objective lenses being employed in the object and the reference arm are 60×/0.7NA and 10×/0.25NA, respectively. (a) and (b) Correlated object and reference speckle fields. (c) and (d) Corresponding specklogram (for SD: stationary diffuser) and interferogram (for RD: rotating diffuser). (e) and (f) Uncorrelated object and reference speckle fields. The decorrelation between the fields is done by translating and tilting the reference Si-wafer. (g) and (h) Corresponding specklogram (for SD: stationary diffuser) and fringe free uniform intensity (for RD: rotating diffuser). The zoomed views of the regions marked with green, red and blue color boxes in both correlated and uncorrelated speckle fields (object and reference) are illustrated in the second and third rows of Fig. 7.

### 4.2. Superposition of the object and the reference field as a function of different magnification

Next, four different objective lenses 10×/0.25NA, 20×/0.45NA, 60×/0.7NA and 60×/1.2NA are employed in the object arm sequentially to understand their effect on the resultant intensity pattern at the camera. The reference arm objective lens is not changed and always kept 10×/0.25NA during all the experimentation. Again, a piece of Si-wafer is used as a sample under the QPM system. First, the diffuser is kept stationary to generate object and reference speckle fields by employing the experimental scheme exhibited in Fig. 6. As it is mentioned in the experimental section, the reference mirror/Si-wafer is attached to a three-axis precision control stage: one translation and two angular motions. The reference mirror is adjusted in a way such that both the arms: object and reference, generate correlated speckle fields at the detector. The reference mirror adjustment is done only with the translation stage to obtain specklograms. The angular motions of mirror controlled the fringe density in the camera FOV as explained in the experimental section.

Figures 8(a) – 8(d) exhibit the speckle fields in the object and the reference arm of the interferometer and their superposition at the camera for all four objective lenses 10×/0.25NA, 20×/0.45NA, 60×/0.7NA and 60×/1.2NA. Figures 8($a_{1-2}$), 8($b_{1-2}$), 8($c_{1-2}$) and 8($d_{1-2}$) illustrate the speckle fields corresponding to the reference and the object arm for 10×/0.25NA, 20×/0.45NA, 60×/0.7NA and 60×/1.2NA objective lenses and represented as R and O. Here, R and O are stand for reference and object arm of the interferometer. The object and the reference arm of the interferometer are sequentially opened in order to record the related speckle fields. The correlation between both the speckle fields for all four objective lenses are calculated using a MATLAB code and listed in Table 1. Figures 8($a_3$), 8($b_3$), 8($c_3$) and 8($d_3$) exhibit their 2D normalized correlation maps, respectively. Further, both the object and the reference arm of the interferometer are opened to generate the specklogram corresponding to all objective lenses as illustrated in Figs. 8($a_4$), 8($b_4$), 8($c_4$) and 8($d_4$), respectively.

It can be clearly visualized that the correlations between the object and the reference arm for all four objective lenses are quite high and leads to the formation of high contrast interference fringes. In order to understand the theory behind obtaining high correlation between both the speckle fields for all four objective lenses in the object arm with a fixed objective lens in the reference arm, first, the average speckle sizes of the speckle fields coming from both the arms are measured. Table 1 exhibits the average speckle sizes of the patterns shown in Figs. 8($a_2$), 8($b_2$), 8($c_2$) and 8($d_2$) for all object arm objective lenses. It can be seen that the average speckle size is measured to be approximately equal for all lenses. This became possible due to the reflection geometry of the present optical configuration, where the speckle field originated from output port of MMF (Fig. 6) is passed twice through both the object and the reference arm objective lenses. The objective lenses, first, demagnified the input speckle field and formed diffraction limited pattern at the sample and the reference mirror. Due to the reflection geometry, the demagnified speckle fields in both the arms are collected by the same objective lens and magnified by the similar amount and speckle size remain unchanged. Thus, the correlation between the speckle fields do not destroy while passing through different objective lenses in the object arm and fixed objective lens in the reference arm and enable the use of non-identical objective lenses in both the arms of the interference setup.

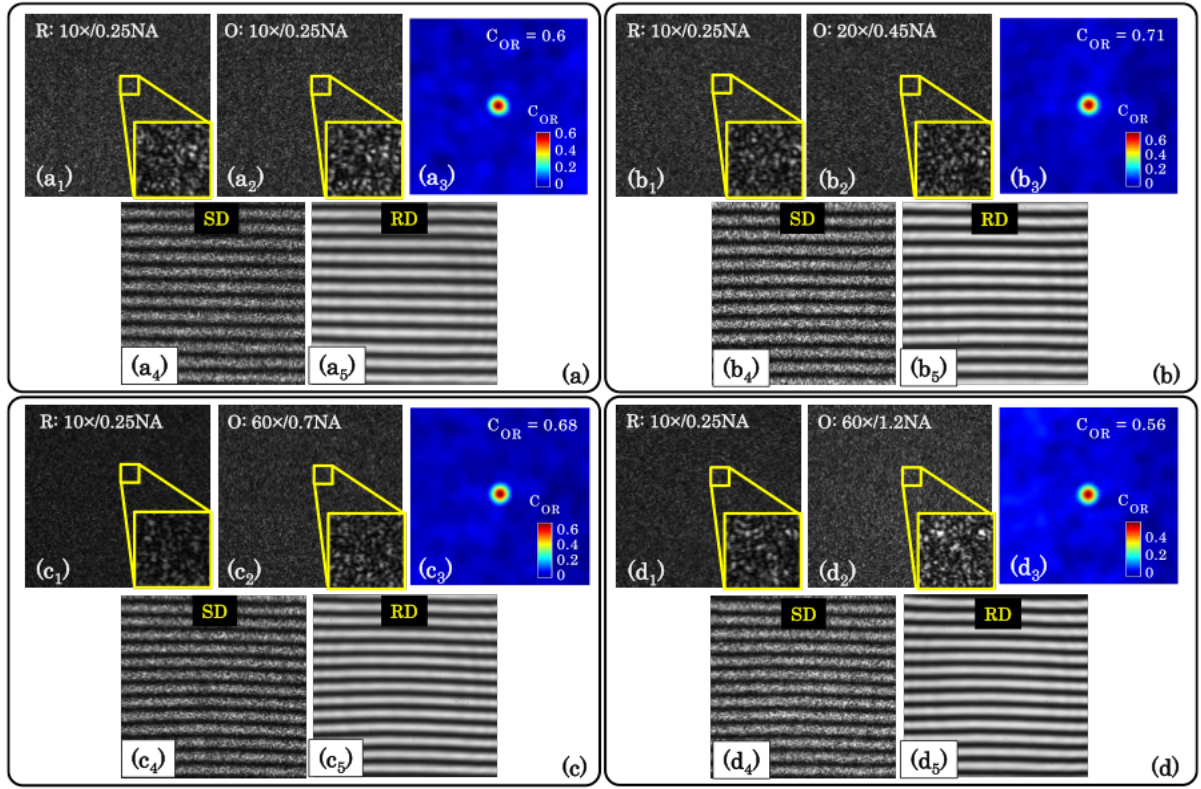

**Fig. 8. Correlation between the object and reference speckle fields as a function of magnification and numerical aperture.** Four different objective lenses 10×/0.25NA, 20×/0.45NA, 60×/0.7NA and 60×/1.2NA are sequentially inserted in the object arm keeping fixed reference arm objective lens 10×/0.25NA. ($a_{1-2}$), ($b_{1-2}$), ($c_{1-2}$) and ($d_{1-2}$) The reference and the object arm speckle fields corresponding to fixed reference objective lens and all four objective lenses in the object arm, respectively. ($a_3$), ($b_3$), ($c_3$) and ($d_3$) illustrate the 2D normalized correlation map between both the speckle fields. ($a_4$), ($b_4$), ($c_4$) and ($d_4$) Superpositions of the object and the reference arm speckle fields also called specklogram corresponding to all four objective lenses in the object arm. The diffuser is kept stationary. ($a_5$), ($b_5$), ($c_5$) and ($d_5$) The temporally averaged patterns of a large numbers of specklograms within the exposure time of the camera. This is achieved by rotating the diffuser at an adequate speed.

**Table 1: Speckle sizes as a function of objective lens magnification and NA**

| S. No. | Objective lens | | Average speckle size (pixels) | $C_{OR}$ (a.u.) |
|---|---|---|---|---|
| | Magnification | Numerical aperture (NA) | | |
| 1. | 10× | 0.25 | 8.8 | 0.60 |
| 2. | 20× | 0.45 | 8.9 | 0.71 |
| 3. | 60× | 0.70 | 8.7 | 0.68 |
| 4. | 60× | 1.20 | 8.2 | 0.56 |

In the transmission geometry, the speckle field is passed only once through the fixed objective lens, therefore, the situation is not identical to the reflection geometry. The switching of the objective lens in the object arm keeping fixed objective lens in the reference arm changes the speckle size of the object arm speckle field when exiting the objective lens and destroy the correlation (i.e., mandatory condition to obtain static/stable specklogram) between both the beams and subsequently the interference fringes. In transmission geometry, identical objective lenses must be used in both the arms in order to satisfy the condition of stable specklogram [13]. Moreover, the distance between the objective lenses and the sensor is also crucial and must be equal to maintain the high correlation between the object and the reference arm speckle fields. If the aforementioned conditions fail to satisfy, it leads to a significant drop in the correlation between the superimposed speckle fields and consequently the fringe contrast of the resultant pattern. Thus, this imposes lots of restrictions for obtaining correlated speckle fields and limits the implementation of QPM system only for a single objective lens.

The specklograms illustrated in Figs. 8($a_4$), 8($b_4$), 8($c_4$) and 8($d_4$) are filled with a speckle noise. The stationary diffuser is then rotated to generate large number specklograms within the exposure time of the camera. The camera exposure time is set to 30 fps. The rotating diffuser generates uncorrelated speckle patterns being used to form specklograms. The average images of large number of specklograms corresponding to all four objective lenses 10×/0.25NA, 20×/0.45NA, 60×/0.7NA and 60×/1.2NA are presented in Figs. 8($a_5$), 8($b_5$), 8($c_5$) and 8($d_5$), respectively. It can be seen that the average images called interferogram are free from the speckle noise and coherent noise. Note that the temporally varying speckle field arises due to rotating diffuser does not the affect the shape and quality of the resultant intensity patterns at the detector for all four objective lenses.

### 4.3. QPM with scalable FOV and resolution

In the previous section, it has been successfully exhibited that different magnification and numerical aperture objective lenses can used in the object arm of the QPM system without affecting the quality of the resultant interference pattern. The fringe density of the interference pattern is kept very low to clearly show the potential of the DSI in QPM system. However, this is not only limited to low fringe density of the interference pattern. The fringe density can be made very high satisfying the Nyquist criteria to implement single shot QPM. The fringe density can be increased by tilting the reference mirror. The tilting of the mirror decorrelates the object and reference speckle fields, which can be adjusted by translating the mirror either towards or away from $MO_1$. The high fringe density interferogram of the USAF chart and its single shot recovered phase map is illustrated in Supplementary Fig. S8.

As DSI has high temporal coherence length. Therefore, to understand the effect of the high temporal coherence length of DSI, identical 10×/0.25NA objective lenses are used in both object and reference arms. The reference mirror and $MO_1$ unit is then translated away from the zero OPD position and subsequently interference patterns are recorded using camera. The non-zero OPD position only changes the shape of fringes and generates circular fringes at the camera without affecting their fringe contrast as illustrated in Supplementary Fig. S6. The correlation between the object and the reference speckle fields as a function of OPD is given in the Supplementary Table S1.

Next, low fringe density phase shifted interferograms of USAF resolution test target (Thorlabs: part # R3L3S1N) are recorded by the camera for the phase recovery. The phase shift between the interferograms is introduced by shifting the reference mirror using piezo transducer with nanometer precision. Here, random phase shifted interferograms are recorded and then PCA based reconstruction algorithm is implemented for the phase recovery of the USAF chart [20]. Figures 9(a) – 9(c) present one of the phase shifted interferograms of the USAF chart corresponding to 10×/0.25NA, 20×/0.45NA and 60×/0.7NA objective lenses, respectively. It can be seen that interference fringes are not perfectly straight in the sample free region which is due to the presence of minute optical aberrations in the QPM system. This leads to the formation of slowly varying background phase in the recovered phase maps, which is numerically compensated. The recovered phase maps of the resolution chart corresponding to 10×/0.25NA, 20×/0.45NA and 60×/0.7NA objective lenses are illustrated in Figs. 9(d) – 9(f), respectively. It can be visualized that the recovered phase maps do not suffer from the speckle noise and coherent noise. Thus, DSI generates highly spatial phase sensitive phase images of the specimens over the whole camera FOV. In addition, it is demonstrated that DSI provides the possibility of obtaining scalable FOV and resolution in QPM system. It is worth noting that the shape of the fringes is not changed while switching different objective lenses in the object arm.

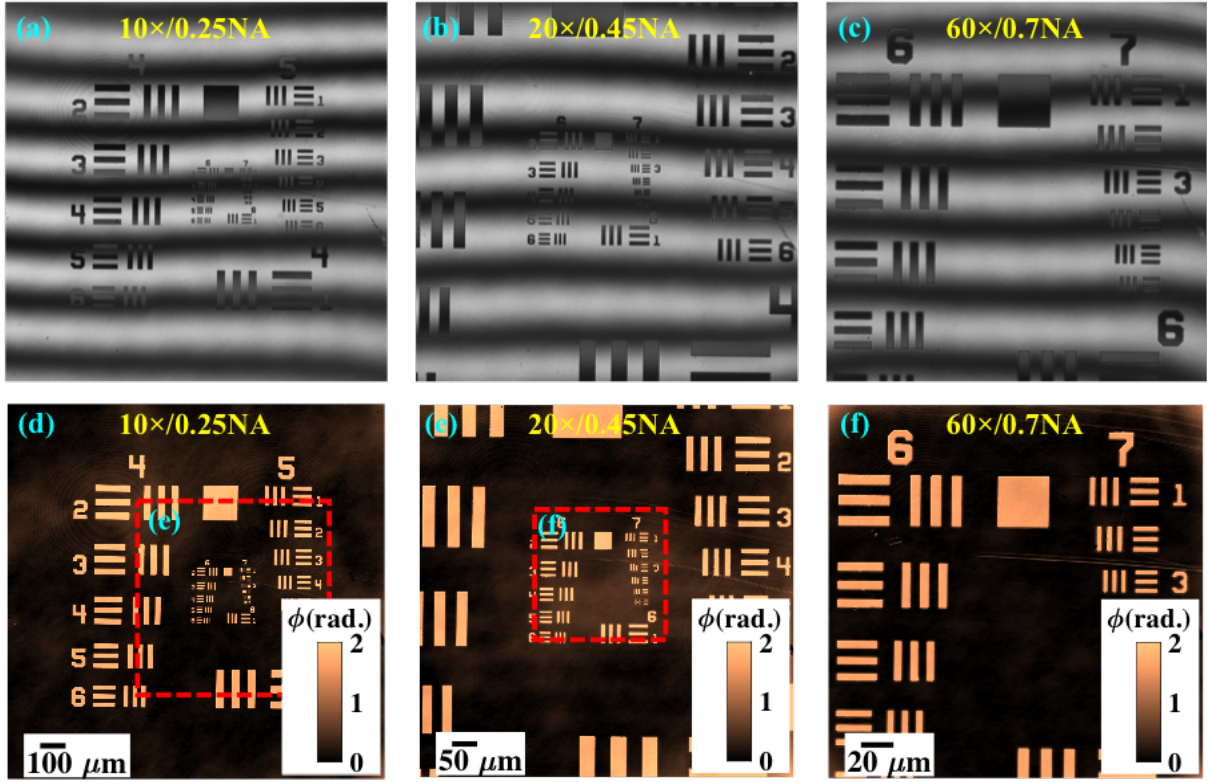

**Fig. 9. DSI provides scalable FOV and resolution.** DSI generates scalable FOV and resolution quantitative phase images without changing the reference objective lens. Three objective lenses with different magnification and numerical aperture were used: 10×/0.25NA, 20×/0.45NA and 60×/0.7NA. (a – c) Interferometric images of USAF resolution chart while using 10×/0.25NA, 20×/0.45NA and 60×/0.7NA objective lens in the object arm, respectively. The insets depict the zoomed view of the region marked with red dotted box. (d – f) Reconstructed phase images of the resolution chart corresponding to aforementioned objective lenses. High temporal coherence of DSI is responsible for the generation of superior quality phase images over the whole camera FOV while employing different magnification and NA objective lenses. The scale bars are in micrometer.

## 5. Discussion and conclusion

In the present work, systematic simulation and experimental studies are done to understand physics behind achieving the interference in dynamic speckle illumination based QPM system. It is observed that DSI forms stable interference signal under rotation of the diffuser if the speckle fields being overlapped are correlated with each other. On the contrary, the superposition of uncorrelated speckle fields does not form nicely oriented fringes in the resultant specklogram and washes out the interference pattern when the diffuser is rotated.

Further, the superposition of a speckle field and its identical spatially shifted version is done to understand its effect on the resultant intensity pattern. It is found that stable interference fringes are formed only if the shift between the speckle fields being superimposed to each other is smaller than the average speckle size. It is observed that high to moderate contrast (i.e., 0.9 to 0.5) interference fringes are formed for the shift less than or equal to half of the speckle size. Interestingly, speckle fields having bigger speckle sizes always exhibited little correlation and lead to non-zero correlation between them.

DSI has high TC length almost equal to the TC length of the parent laser light source and low SC length depending on the source size. High TC length of DSI helps to achieve the interference pattern quickly in QPM systems compared to low TC light sources such as halogen lamp, light emitting diodes etc. DSI can also be used to balance the object and the reference arm of reflection QPM systems, to obtain interference patterns with low TC light sources. In addition, dynamic speckle illumination derived from a high TC length light source enables the user defined magnification and resolution in the QPM system. DSI provides high spatial phase sensitivity comparable to low TC length light source in QPM system [15]. Most of the biological applications require a highly spatial phase sensitive QPM phase microscope with the possibility of changing objective lenses of different magnifications and numerical apertures in the object arm. This would make DSI-QPM system a suitable candidate for several biological applications. Here, it is important to highlight that changing the objective lens only in the object arm disturbs the zero OPD between the object and the reference beam. As the OPD between both the beams must be smaller than the TC length of light source to observe interference pattern at the detector. Therefore, low TC light sources cannot form interference fringes in such unbalanced interferometric system. On the contrary, DSI being high TC in nature can form the interference fringes even in an unbalanced QPM configuration provided both the fields reaching at the detector are correlated and unshifted. Moreover, the photon degeneracy of DSI is very high compared to halogen lamp or LEDs [21, 22]. The high photon degeneracy can also be obtained by employing direct narrow band laser at the cost of unwanted interference fringes due to the coherent superposition of multiple reflections coming from surfaces of the optical components [13, 15, 23]. As a consequence, it reduces the spatial phase sensitivity and measurement accuracy of the system.

Thanks to the reflection geometry of QPM employed with DSI, which provides the freedom of moderate translation (several mm) of the reference arm (mirror and objective lens both) without affecting the contrast of the interference fringes (see Supplementary Fig. S6). Thus, the precise positioning of the object and the reference arm within few tens of micrometers is not required as in the case of low TC light source. DSI-QPM generates straight, circular and curved fringes depending on the position of the reference arm (see Supplementary Fig. S6). In addition, the fringe density can be easily tuned from extremely low to extremely high over the whole camera FOV without affecting the interference fringe quality (see Supplementary Fig. S8). Thus, DSI enables both single-shot and multi-shot quantitative phase recovery of the specimens under test. In order to support our claim, the scalable FOV with scalable resolution quantitative phase imaging of a standard USAF resolution test target in both single-shot and multi-shot is demonstrated. We believe the present work would provide a deeper understanding of DSI-QPM and its potential impact for different applications.

**Acknowledgments:**

**Funding:**

B.S.A. acknowledges the funding from the INTPART (project # 309802), Research Council of Norway, project NANO 2021–288565 and BIOTEK 2021–285571.

**Author Contributions:**

A.A. has conceptualized the idea, designed and performed most of the experiments, and analyzed the data. A.A. developed the theory and performed the computations. A.A. and B.S.A. mainly wrote the manuscript. N.J. discussed the theory, carried out some experiments and assisted during all the experiments. B.S.A. conceived the project and supervised this work. All authors reviewed and edited the manuscript.

**Competing interests:**

The authors declare no competing interests.

**Data and materials availability:**

The authors declare the availability of the data and codes used in the research to obtain the results reported in the manuscript upon reasonable request.

# Supplementary Materials

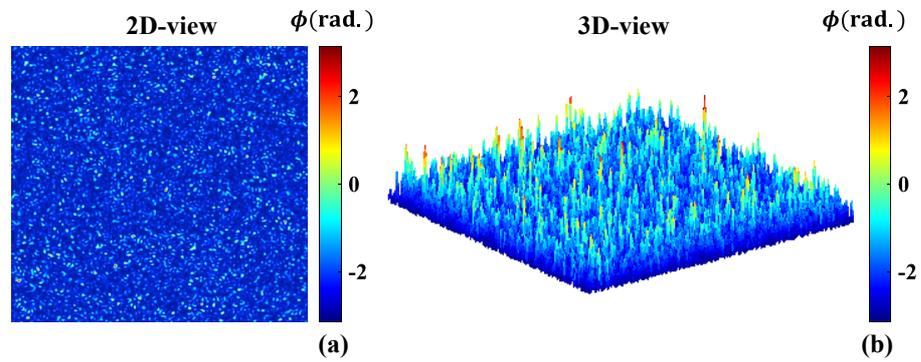

**Fig. S1.** (a) and (b) 2D and 3D views of the simulated rough surface being used to generate speckle fields for the simulation studies. The phase values of the rough surface vary from – pi to + pi.

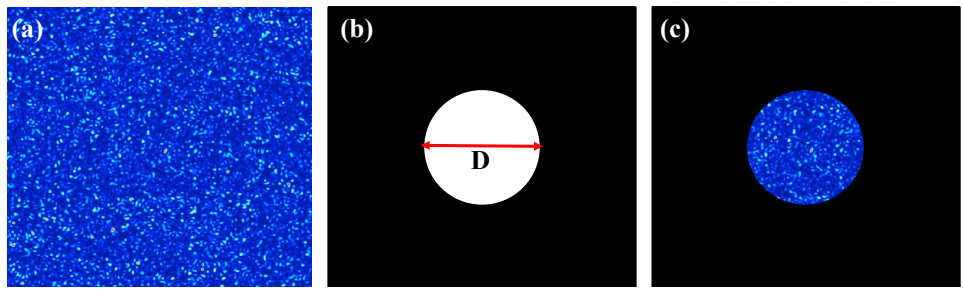

**Fig. S2.** (a) Simulated rough surface being used to generate speckle fields for the simulation studies. (b) Binary mask with a circular opening of diameter 'D'. The diameter of the opening decides the laser beam diameter hitting the rough surface. (c) Multiplication of the rough surface and binary mask to limit the contribution of the scattering sites falling within the opening region of the mask on the resultant speckle pattern. This way it mimics the experimental situation where the scattering sites interacting with the laser only contribute to the resultant speckle field.

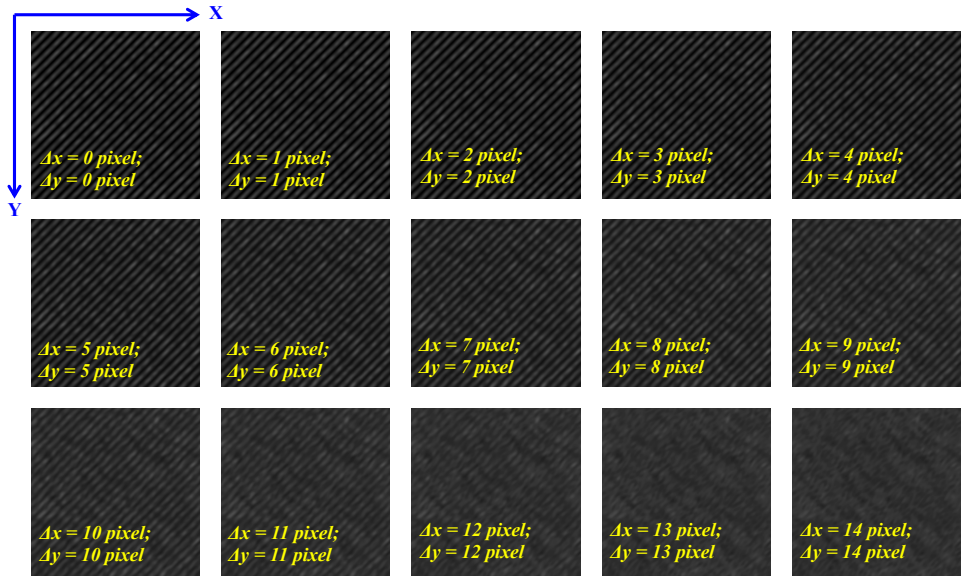

**Fig. S3.** The average images of 360 specklograms for a speckle size of 15 pixel corresponding to different shifts in one of the speckle fields being superimposed with another identical unshifted speckle field. The shift is done sequentially from 0 pixel to the average speckle size in a step of 1 pixel.

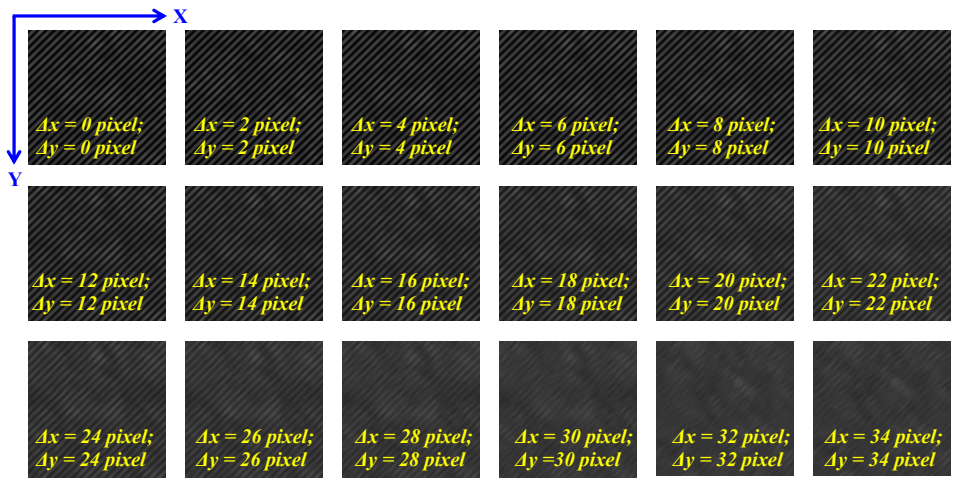

**Fig. S4.** The average images of 360 specklograms for a speckle size of 30 pixel corresponding to different shifts in one of the speckle fields being superimposed with another identical unshifted speckle field. The shift is done sequentially from 0 pixel to the average speckle size in a step of 2 pixel.

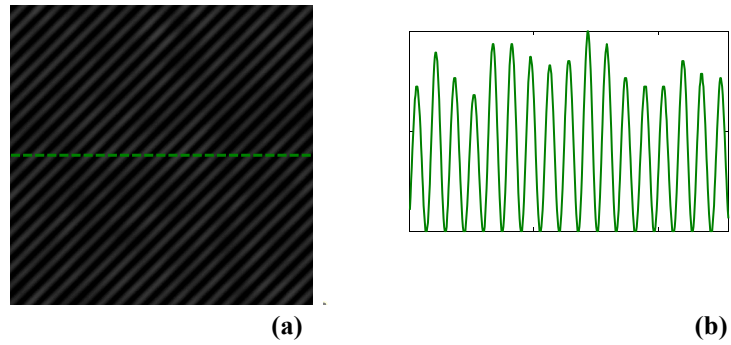

(a)                                                   (b)

**Fig. S5.** (a) Average interferogram of the 360 simulated statistically independent specklograms corresponding to different rotation angles of the diffuser from 1° to 360° in a step of 1°. The specklograms are generated due to the supposition of two correlated speckle fields (see main text Fig. 3 for more details) of speckle size of 30 pixel. (b) The line profile of the average image called interferogram along the green dotted horizontal line. It can be seen that modulation depth of the interferogram is not constant over the entire FOV. This could be due to the insufficient averaging of the statistically independent specklograms. This can be improved by generating large number of statistically independent specklograms.

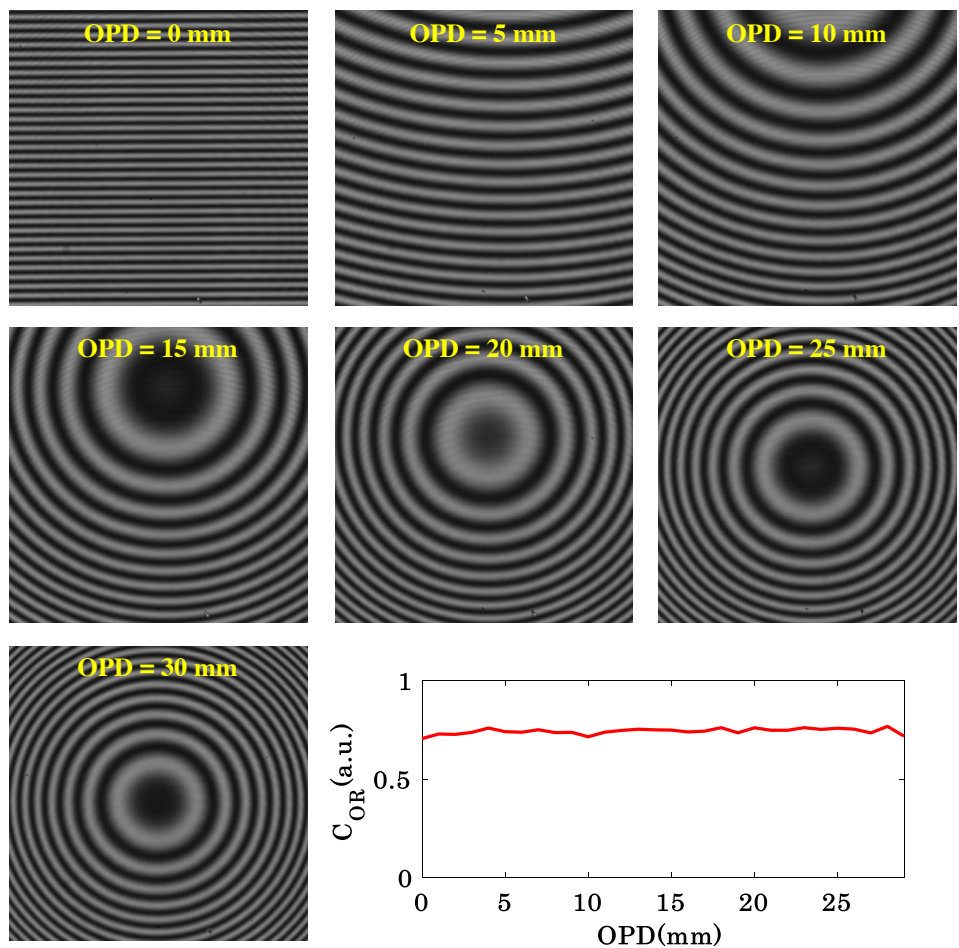

**Fig. S6.** The interferograms generated due to the superposition of correlated speckle fields in both the object and the reference arms of the interferometer as a function of the optical path difference (OPD) between them. For this study, identical 10×/0.25NA objective lenses are used in both the object and the reference arms of QPM. The reference arm unit (objective lens MO2 and mirror M together) is translated to introduce OPD between the object and the reference

arm. It can be visualized that non-zero OPD does not washout the interference pattern due to the high temporal coherence length of PTLS. The non-zero OPD only generates curved or circular fringes at the camera. The OPD is increased from 0 mm to 35 mm in a step of 1 mm. The line profile illustrates the variation of the normalized correlation between the object and the reference speckle fields as a function OPD. It can be seen that the correlation does not vary as a function of OPD between the speckle fields. Thanks to the reflection geometry of QPM, The size and the position of the speckles do not change in the reference speckle field. This is due to the double pass of the speckle field through the reference arm objective lens which cancels the effect of the speckle field propagation on speckle's size and shape.

**Table S1. The normalized correlation between the object and reference arm speckle fields as a function of OPD between them.**

| OPD (mm) | $C_{OR}$ | OPD (mm) | $C_{OR}$ | OPD (mm) | $C_{OR}$ | OPD (mm) | $C_{OR}$ | OPD (mm) | $C_{OR}$ |
|---|---|---|---|---|---|---|---|---|---|
| 0 | 0.70 | 6 | 0.73 | 12 | 0.74 | 18 | 0.76 | 24 | 0.75 |
| 1 | 0.72 | 7 | 0.75 | 13 | 0.75 | 19 | 0.73 | 25 | 0.75 |
| 2 | 0.72 | 8 | 0.73 | 14 | 0.75 | 20 | 0.76 | 26 | 0.75 |
| 3 | 0.73 | 9 | 0.73 | 15 | 0.74 | 21 | 0.74 | 27 | 0.73 |
| 4 | 0.75 | 10 | 0.71 | 16 | 0.73 | 22 | 0.74 | 28 | 0.76 |
| 5 | 0.74 | 11 | 0.73 | 17 | 0.74 | 23 | 0.76 | 29 | 0.71 |

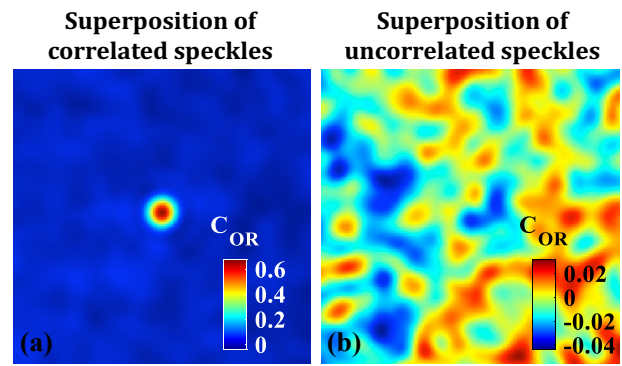

**Fig. S7.** (a) and (b) Normalized correlations maps corresponding to the experimentally recorded correlated and uncorrelated speckle patterns being used to generate specklograms. Fig. S7(b) does not show any correlation peak.

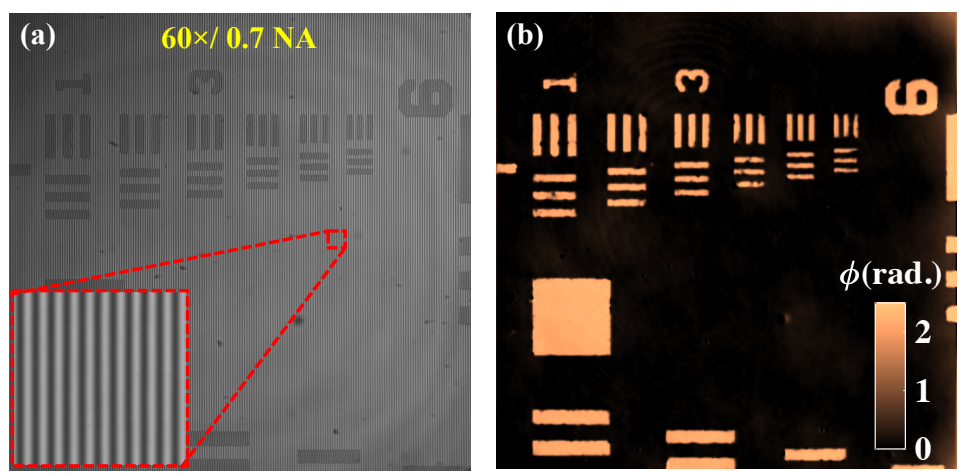

**Fig. S8.** (a) High fringe density interferometric image of USAF resolution chart while using 10×/0.25NA in the reference arm and 60×/0.7NA objective lens in the object arm. The insets depict the zoomed view of the region marked with red dotted box. (b) Single-shot recovered phase image of the resolution chart corresponding to aforementioned objective lenses. The color bar is in rad.